\documentclass[11pt]{article}
\usepackage{amsmath,amssymb,color,epsfig,cite}
\usepackage{graphicx}
\usepackage{subfigure}
\usepackage{setspace}

\usepackage{amsfonts}

\newcommand{\hoch}[1]{$\, ^{#1}$}

\makeatletter
\@addtoreset{equation}{section}
\makeatother

\newcommand{\be}{\begin{equation}}
\newcommand{\ee}{\end{equation}}
\newcommand{\bea}{\setlength\arraycolsep{2pt} \begin{eqnarray}}
\newcommand{\eea}{\end{eqnarray}}
\newcommand{\nn}{\nonumber}

\def\ft#1#2{{\textstyle{\frac{\scriptstyle #1}{\scriptstyle #2} } }}
\def\fft#1#2{{\frac{#1}{#2}}}

\def\0{{\sst{(0)}}}
\def\1{{\sst{(1)}}}
\def\2{{\sst{(2)}}}
\def\3{{\sst{(3)}}}
\def\4{{\sst{(4)}}}
\def\5{{\sst{(5)}}}
\def\6{{\sst{(6)}}}
\def\7{{\sst{(7)}}}
\def\8{{\sst{(8)}}}
\def\sst#1{{\scriptscriptstyle #1}}

\def\ep{{\epsilon}}

\def\im{{{\rm i\,}}}

\def\Dslash{\slash \negthinspace \negthinspace \negthinspace \negthinspace D}

\begin{document}

\begin{flushright}
\hfill { MI-HET-794
}\\
\end{flushright}

\begin{center}
{\large {\bf De-Higgsing In Eleven-Dimensional Supergravity On 
The Squashed $S^7$
 }}

\vspace{15pt}
{\large B.E.W. Nilsson$^1$ and 
              C.N. Pope$^{2,3}$}

\vspace{15pt}

\hoch{1}{\it Department of Physics, Chalmers University of Technology,
 SE-412 96 G\"oteborg, Sweden}

\hoch{2}{\it George P. \& Cynthia Woods Mitchell  Institute
for Fundamental Physics and Astronomy,\\
Texas A\&M University, College Station, TX 77843, USA}

\hoch{3}{\it DAMTP, Centre for Mathematical Sciences,
 Cambridge University,\\  Wilberforce Road, Cambridge CB3 OWA, UK}

\vspace{20pt}



\end{center}




\begin{abstract}

In this paper we construct the subset of modes on $S^7$ that are 
relevant in the 
compactification of eleven-dimensional supergravity on a squashed $S^7$
when restricted to the sector that comprises singlets under the 
$Sp(1)\times Sp(2)$ isometry of the squashed sphere.  
Some of the properties of these modes, connected to  the transition from 
the round $S^7$  to the squashed $S^7$, are analysed in detail.  
Special features of the Rarita-Schwinger operator,
described in earlier work by Buchdahl, are explained and related to 
properties of the squashed $S^7$ operator spectrum  obtained in 
previous work by the authors. 
We then discuss how  the singlet modes give rise to 
supermultiplets in the left-squashed case, the phenomenon of de-Higgsing,  
and what happens to the AdS$_4$ fields in these supermultiplets under 
an orientation reversal (``skew-whiffing'') of the squashed $S^7$. 
Finally, we  consider the 
possible choices of boundary conditions that appear for some of these 
fields in AdS$_4$ in the case of the right-squashed non-supersymmetric 
compactification, and 
how these choices may affect the stability of the gravity theory.

\end{abstract}

\vfill
{\scriptsize
bengt.ew.nilsson@chalmers.se, \ pope@physics.tamu.edu}
\pagebreak

\tableofcontents
\addtocontents{toc}{\protect\setcounter{tocdepth}{2}}

\section{Introduction}

 There has recently been a lot of effort put into deriving the complete 
spectrum of some non-supersymmetric AdS vacua of maximal supergravity theories 
in ten and eleven dimensions.
One reason for this endeavour is to challenge, or vindicate, conjectures 
related to the swampland project,
in particular the AdS stability conjecture proposed by Ooguri and 
Vafa \cite{Ooguri:2016pdq}.

This issue becomes especially interesting when discussing vacua that are 
BF (Breitenl\"ohner-Freedman) stable but which so far have not been proven 
unstable despite
having no supersymmetry. In fact, for $D=11$ supergravity, the squashed 
seven-sphere vacuum without supersymmetry,
the so called right-squashed sphere \cite{dunipo0, dunipo}, is one of a very small 
set of such examples\footnote{Another case suggested recently is based on a deformed S-fold compactification \cite{Guarino:2022tlw, Guarino:2020flh}.}.
The squashed $S^7$ spectrum of fields and their $SO(2,3)$ irrep content 
has recently been derived in full 
detail \cite{Nilsson:2018lof, Ekhammar:2021gsg, Karlsson:2021oxd, Karlsson-Nilsson:2023}\footnote{The spectrum of the squashed seven-sphere was reproduced recently in \cite{Duboeuf:2022mam} by entirely different methods.}
which makes it possible  to study also the relation between the round 
$S^7$ compactification and its cousins
with partially or totally broken supersymmetries, i.e., the left-squashed 
and right-squashed $S^7$ vacua with $\mathcal{N}=1 $ and 
$\mathcal{N}=0 $ supersymmetry, respectively. Some aspects of this relation for scalar and spin-$\ft12$ modes
were addressed in \cite{Nilsson:1983ru}.

A number of comments on this relation were made in 
\cite{Nilsson:2018lof}. Here we discuss an additional aspect connected 
to, in particular, the Rarita-Schwinger operator and properties of its 
eigenvalue equation on non-Einstein versions of the squashed $S^7$.
This is, in fact, a well-known issue which was studied some time ago by 
Buchdahl  \cite{Buchdahl:1958xv, Buchdahl:1962, Buchdahl:1982ni}. 
In the present context it becomes especially intriguing 
since for a large portion
of the mode spectrum on $S^7$ can be traced continuously between the round and 
the Einstein-squashed vacuum solutions of $D=11$ supergravity (see \cite{Nilsson:1983ru}), 
while Buchdahl's
 results indicate that this should not be possible in general.
 
Our results are based on an explicit construction of
the entire set of singlet mode functions on the squashed $S^7$.  Although
they constitute only a small subset of the totality of modes in the complete
spectrum, the singlet modes are of particular interest because they 
form a consistent truncation in their own right, and additionally, they
have elegant geometrical interpretations, allowing their simple explicit
construction.  They also exhibit some of the most intriguing features in the
relation between the round and the squashed $S^7$ vacua.
The complete singlet sector involves in addition to the singlet 
Rarita-Schwinger mode relevant for the Buchdahl issue
also its superpartners, the  ``squashing mode,''  which is the Lichnerowicz operator eigenmode  first discussed
 by Page in \cite{page}, and two singlet modes of the first-order
3-form operator $Q={*d}$.  
Adding to these the scalar and spin-$\ft12$ 
singlet modes 
 they make up precisely two Wess-Zumino supermultiplets. The latter 
two singlet modes are also responsible for 
the ${\mathcal N}=1$ supergravity multiplet.
 The properties of these modes on the non-supersymmetric right-squashed $S^7$ are also analysed in detail.

The paper is organised as follows. In section 2  we give some basic material needed in the following sections. Section 3 contains the construction of the singlet modes on $S^7$ and  
a discussion of some of their properties connected to  the transition from the round $S^7$ compactification to the squashed one in particular the special features of the Rarita-Schwinger operator. 
Then, in section 4, we explain how the singlet modes give rise to 
supermultiplets in the left-squashed case and what happens to the AdS$_4$ fields in these supermultiplets under skew-whiffing. 
The 
possible choices of boundary conditions in the right-squashed 
non-supersymmetric case are identified and  possible  implications for 
the issue of stability are discussed.
The final section contains a summary and some conclusions. 

\section{Preliminaries}

We recall that the mass operators for the various towers in the
spectrum of AdS$_4$ supergravity from compactification on $M_7$ are given by
 (see table 5 of Phys. Reps. \cite{dunipo}\footnote{We have changed  to standard conventions where $\{\gamma_a, \gamma_b\} =
 + 2\delta_{ab}$, by sending the $\Gamma_a$ Dirac matrices in \cite{dunipo} 
to $-\im \gamma_a$.})
\bea
2^+ \qquad\qquad M^2&=& \Delta_0\nn\\
\ft32^{\1, \2} \qquad\qquad M&=& -\im \,\Dslash_{1/2} + \ft72 m\nn\\
1^{-\, \1, \2} \qquad\qquad M^2 &=& \Delta_1 + 
  12 m^2 \pm 6m (\Delta_1 + 4 m^2)^{\ft12}\nn\\
1^{+} \qquad\qquad M^2 &=&\Delta_2\nn\\
\ft12^{\4, \1} \qquad\qquad M&=& -\im\, \Dslash_{1/2} -\ft92 m\nn\\
\ft12^{\3, \2}\qquad\qquad M &=& \ft32 m + \im\, \Dslash_{3/2}\nn\\
0^{+\, \1, \3} \qquad\qquad M^2 &=& \Delta_0 + 
     44m^2 \pm 12 m (\Delta_0 + 9 m^2)^{\ft12}\nn\\
0^{+\, \2} \qquad\qquad M^2 &=& \Delta_L - 4m^2\nn\\
0^{-\, \1, \2} \qquad\qquad M^2&=& (Q+3m)^2 - m^2\,.\label{massops}
\eea
  The internal space $M_7$ carries an Einstein 
metric, with Ricci tensor given by $R_{ab}=6m^2\, g_{ab}$. 
The remaining notation here is as follows:

 The entries in the left-hand column in (\ref{massops}) denote the spins
of the four-dimensional fields in the Kaluza-Klein towers.  The 
notation of the additional superscripts is explained in detail in
\cite{dunipo}.  The operators appearing in the right-hand column
are defined as follows:
\bea
\hbox{Scalar Laplacian}: && \Delta_0\, \phi=-\square\,\phi\,,\nn\\
\hbox{Vector Hodge-de Rham}:&& \Delta_1 \,V_a= -\square\, V_a +
   R_a{}^b\, V_b\,,\nn\\
\hbox{Lichnerowicz operator}:&& \Delta_L\,h_{ab}=  
-\square\, h_{ab} - 2R_{acbd}\, h^{cd} + 2R_{(a}{}^c\, h_{b)c}\,,\nn\\
\hbox{2-form Hodge-de Rham}:&& \Delta_2 \,\omega_{ab}=-\square\,\omega_{ab} 
  -2 R_{acbd}\, \omega^{cd} -2 R^c{}_{[a}\, \omega_{b]c}\,,\nn\\
\hbox{3-form operator}:&& Q\,\omega_3= {*d}\, \omega_3\,,\nn\\
\hbox{Dirac operator}:&& \Dslash_{1/2}\,\psi= \gamma^a\, D_a\,\psi\,,\nn\\
\hbox{Rarita-Schwinger operator}:&& \gamma^{abc}\nabla_b\psi_c\,.
\eea
For vectors, the transversality condition $\nabla_a V^a=0$ is imposed and similarly for 
2-forms and 3-forms. Lichnerowicz
modes are required to be transverse and traceless and  modes of 
the Rarita-Schwinger operator are required to be gamma-traceless,
$\gamma^a\psi_a=0$.  As will be discussed later, consistency also
then implies that $\nabla_a\psi^a=0$ and that the metric be Einstein.
The Rarita-Schwinger operator then reduces to the Dirac operator
acting on vector-spinors, which we denote by $\Dslash_{3/2}$.

  We shall be concerned with the spectrum of modes in the round and
squashed $S^7$ vacua of $D=11$ supergravity.  
The family of squashed $S^7$ metrics, described as an $SU(2)$ bundle 
over $S^4$, is given by
\be
ds^2 = d\mu^2 +\ft14 \sin^2\mu\, \Sigma_i^2 + \lambda^2\, (\sigma_i- A^i)^2\,.
\label{squashedS7}
\ee
Here $A^i= \cos^2\ft12 u\, \Sigma_i$, and $\sigma_i$ and $\Sigma_i$ are
two sets of left-invariant 1-forms on the group manifold $SU(2)$.  They
obey
\bea
d\sigma_i= -\ft12 \ep_{ijk}\, \sigma_k\,,\qquad
d\Sigma_i=-\ft12 \ep_{ijk}\, \Sigma_k\,.
\eea
When
required, we use the vielbein basis $e^a$ with
\be
e^0=d\mu\,, \qquad e^i= \ft12 \sin\mu\, \Sigma_i\,,\qquad
e^{\hat i}= \lambda\, \nu_i=\lambda\, (\sigma_i- A^i) \,,\label{vielbein}
\ee
where $i$ ranges over the values 1, 2 and 3, and $\hat i$= ranges over
4, 5 and 6, with $\hat1 =4$, $\hat 2=5$ and $\hat 3=6$.  The constant
$\lambda$ is the squashing parameter.  The isometry group of the
metric is $Sp(1)\times Sp(2)$ (which is locally the same as
$SU(2)\times SO(5)$) for generic values of $\lambda$, 
enhancing to $SO(8)$ in the case $\lambda=1$, which corresponds to
the round $S^7$.

The metric (\ref{squashedS7}) is Einstein when $\lambda=1$, obeying
\bea
R_{ab}= 6 g_{ab}\,,\label{round}
\eea
and when $\lambda=1{\sqrt5}$, with
\bea
R_{ab}= \fft{27}{10} g_{ab}\,\label{esquashed}\,.
\eea
For general values of $\lambda$, the Ricci scalar is
\be
R= \fft{3(1+8\lambda^2-2\lambda^4)}{2\lambda^2}\,.
\ee
It will be convenient, when calculating the eigenvalues, to rescale the metric
so that it has $R=42m^2$ for all values of the squashing parameter $\lambda$.
This will imply that eigenvalues of a linear operator will  need to be
scaled by the factor
\be
\fft{2\sqrt 7\, \lambda\, m}{\sqrt{1+8\lambda^2 - 2\lambda^4}}\,.\label{norm}
\ee
Eigenvalues of a second-order operator will need to be scaled by
the square of (\ref{norm}). Note that after the rescaling of the metric,
the round and the squashed Einstein metrics will both depend on the parameter $m^2$ and obey 
\bea
R_{ab}=6m^2\, g_{ab}\,.
\eea

\section{Singlet Modes On The Squashed $S^7$}

  In this section, we shall collect together a number of previously-known
results for singlet modes of the relevant operators on the squashed $S^7$
whose eigenvalues govern the masses of the corresponding four-dimensional
fields in the $S^7$ compactifications of $D=11$ supergravity. Some of
our discussion here will also extend beyond that previously presented in
the literature.  The results we obtain here will then be used in the
next section in order to study the mass spectrum in the singlet sector
for both the left-squashed and the right-squashed $S^7$ vacua, and to
relate these spectra to the spectrum in the round-sphere vacuum.

The singlet modes we shall consider in this section comprise the
singlet mode of the scalar Laplacian $\Delta_0=-\square$ (this, of course, is 
just the trivial constant mode); a singlet in the spectrum of the
Lichnerowicz operator
$\Delta_L$ acting on transverse, traceless 2-index symmetric tensors; 
two  singlets in the spectrum of the first-order operator $Q={*d}$ acting on
3-forms; a singlet in the spectrum of the Dirac operator acting on 
spin-$\ft12$ fields on the $S^7$; and a singlet
in spectrum of the Rarita-Schwinger (RS) operator acting on transverse
spin-$\ft32$ fields on the $S^7$.  In all, these comprise the entire
set of singlet modes that are relevant for determining the spectrum of
fields in the $S^7$ compactifications of $D=11$ supergravity.  Note that
the Hodge-de Rham operators $\Delta_1$ and $\Delta_2$ acting on 1-forms
and 2-forms have no $Sp(1)\times Sp(2)$ singlet modes in 
the round or squashed $S^7$ backgrounds.

\subsection{Scalar singlet modes}

The complete spectrum of the scalar Laplacian $\Delta_0=-\square$ on
the whole family of squashed $S^7$ metrics was obtained in 
\cite{Nilsson:1983ru}.  For our present purposes, it suffices to
note that the only singlet in the spectrum of $\Delta_0$ is the 
constant mode, with eigenvalue 0.

\subsection{Lichnerowicz singlet modes}

 There is only one singlet in the spectrum of the Lichnerowicz operator
acting on transverse, traceless symmetric tensor $h_{ab}$ in
the squashed $S^7$ metric, namely the so-called ``squashing mode'' that
corresponds to an infinitesimal variation of the squashing parameter
while keeping the volume fixed.  In the orthonormal frame defined in
eqn (\ref{vielbein}), it corresponds to a tensor whose non-vanishing
components (up to an constant scale) are given by
\bea
h_{00}=3\,,\qquad h_{ij}=3\delta_{ij}\,,\qquad h_{\hat i\hat j}= 
 -4\delta_{ij}\,.\label{Lichmode}
\eea
It was first discussed in \cite{page}.  A straightforward calculation
shows that in the background of the metric (\ref{squashedS7}), it
is an eigentensor satisfying $\Delta_L\, h_{ab}=\sigma_L\, h_{ab}$
with $\sigma_L= 7\lambda^2$.  After making the rescaling by two powers
of the factor in eqn (\ref{norm}), so as to normalise to squashed metrics
with $R=42m^2$, we get the Lichnerowicz eigenvalue
\be
\sigma_L= \fft{196\, \lambda^4\, m^2}{1+8\lambda^2-2\lambda^4}\,.
\label{Lichev}
\ee

   It should be noted that because the eigenfunctions of the Lichnerowicz
operator are defined for all values of the squashing, we can in 
particular continuously
follow them as we pass from the round $S^7$ to the Einstein-squashed $S^7$.
For the particular case of the singlet mode (\ref{Lichmode}), we see
from eqn (\ref{Lichev}) that we have
\bea
\lambda=1:&& \sigma_L = 28m^2\,,\nn\\
\lambda=\ft1{\sqrt5}:&& \sigma_L=\fft{28m^2}{9}\,.
\eea
These values accord with those obtained in \cite{page}.  Note that the
Lichnerowicz eigenvalue $\sigma_L = 28m^2$ on the round $S^7$ is 
in accordance with the known round-sphere results, where the lowest 
level of TT Lichnerowicz modes are in the ${\bf 300}$ representation
of $SO(8)$, and have eigenvalue $28m^2$ (see, for example, \cite{dunipo}). 
And indeed, as must be the case, the ${\bf 300}$ of $SO(8)$ has a
singlet in its decomposition under the $Sp(1)\times Sp(2)$ subgroup
(the isometry group of the squashed $S^7$).

  In fact one can straightforwardly see that this $Sp(1)\times Sp(2)$ 
singlet is the unique singlet in the entire spectrum of $SO(8)$
representations for TT eigentensors of the Lichnerowicz operator
on the round sphere.  This justifies the statement made above that
the tensor (\ref{Lichmode}) is the unique singlet TT Lichnerowicz 
mode in the squashed $S^7$ background.\footnote{A crucial point in
the argument is the fact, mentioned above, that every Lichnerowicz 
mode can be continuously followed from the round sphere to any of the
squashed sphere family of metrics.}

\subsection{Singlet modes of the $Q={*d}$ operator on 3-forms}

   3-form modes of the operator $Q={*d}$ exist for all values of the
squashing parameter, and therefore we can again count the number of
singlet modes by looking at the decomposition of the known $SO(8)$
representations for 3-form modes on the round sphere, and counting
the number of singlets in their decomposition under $Sp(1)\times Sp(2)$.
There are in fact two singlets in total.  

   We can construct these singlet modes by considering 3-forms in the
squashed sphere metric (\ref{squashedS7}) of
the form
\be
\omega= \alpha\, F^i\wedge \nu_i + \beta \nu_1\wedge\nu_2\wedge \nu_3\,,
\ee
where $\alpha$ and $\beta$ are constants, $\nu_i\equiv \sigma_i - A^i$, and
$F^i= dA^i +\ft12\epsilon_{ijk}\, A^j\wedge A_k$.
Solving for $Q\omega=\sigma \omega$, one finds
\be
\alpha = \fft{\sigma\, \beta}{6\lambda^3}\,,\qquad
\alpha(1-\sigma\lambda) = -\beta\,,
\ee
which can be solved for $\sigma$ giving
\be
\sigma_\pm= \fft1{2\lambda} \pm \fft1{2\lambda}\, \sqrt{1+24\lambda^4}\,.
\ee
Normalising the squashed sphere so that its Ricci scalar is $R=42m^2$ for
all $\lambda$ implies we should scale these eigenvalues by the factor
given in (\ref{norm}), and so the eigenvalues become
\bea
\sigma_+ &=& \fft{\sqrt7\, m\, [\sqrt{1+24\lambda^4}+1]}{
                     \sqrt{1+8\lambda^2 - 2\lambda^4}}\,,\\
\sigma_- &=& -\fft{\sqrt7\, m\, [\sqrt{1+24\lambda^4}-1]}{
                     \sqrt{1+8\lambda^2 - 2\lambda^4}}\,.
\eea

Thus for the $\sigma_+$ mode we have
\bea
\lambda=1:&& \sigma_+ = 6m\,,\nn\\
\lambda=\ft1{\sqrt5}:&& \sigma_+=4m\,.
\eea
This 3-form mode is the singlet in the
decomposition of the 840-dimensional $(2,0,0,2)$ $SO(8)$ representation
in the round-sphere vacuum.

For the the $\sigma_-$ mode we have
\bea
\lambda=1:&& \sigma_- = -4m\,,\nn\\
\lambda=\ft1{\sqrt5}:&& \sigma_-=  -\fft{2m}{3}\,.
\eea
This 3-form mode is the singlet in the
decomposition of the 35-dimensional $(0,0,2,0)$ $SO(8)$ representation.

  Of course, since $Q={*d}$ is a first-order operator its eigenvalues
will reverse in sign if the orientation of the manifold is reversed.

\subsection{Singlet modes of the Dirac operator}

 Turning now to the fermionic modes, we begin with the Dirac operator
acting on spinor modes.  Modes of the Dirac operator can be followed
continuously from the round to the squashed sphere, and so the singlet
Dirac modes on the squashed sphere can be counted by counting the number
of singlets in the decomposition of the $SO(8)$ Dirac mode representations
of the round sphere under the $Sp(1)\times Sp(2)$ subgroup. There is
in fact a unique such singlet.  These and the following features of the Dirac equation
can be found  in \cite{Nilsson:1983ru}.

   The $Sp(1)\times Sp(2)$ singlet Dirac mode can be constructed by 
first constructing the Killing spinor $\eta$ in the Einstein-squashed
$S^7$.  Using the orthonormal frame in eqn (\ref{vielbein}), and
making a natural choice for the spin frame, $\eta$ is a spinor with
constant components, satisfying
\bea
\nabla_a\eta =-\fft {\im m}{2}\, \gamma_a\eta\label{KSeta}
\eea
in the $\lambda=\ft1{\sqrt5}$ squashed $S^7$ (we have rescaled the metric so
that $R=42m^2$, as discussed previously).  A straightforward calculation
shows that the same constant-component spinor $\eta$ is an eigenspinor of the
Dirac operator for arbitrary values of the squashing parameter, and
its eigenvalue $\sigma$, defined by $\im\Dslash_{1/2}\, 
\eta = \sigma\,\eta$, is
\bea
\sigma = \fft{3\sqrt7 (1+2\lambda^2)\, m}{2\sqrt{1+8\lambda^2-2\lambda^4}}
\,.
\eea
Thus in particular, on the round $S^7$ and on the Einstein-squashed
$S^7$ we have
\bea
\lambda=1:&& \sigma = \fft{9m}{2}\,,\nn\\
\lambda=\ft1{\sqrt5}:&& \sigma= \fft{7m}{2}\,.
\eea
This Dirac mode is the singlet in the decomposition of the 56-dimensional
$(1,0,0,1)$ $SO(8)$ representation of Dirac modes on the round $S^7$.

\subsection{Singlet modes of the Rarita-Schwinger operator}

   Consider a Rarita-Schwinger mode obeying 
\be
\im\gamma^{abc}\,\nabla_b\psi_c= \sigma\, \psi^a\,.\label{RSop}
\ee
Multiplying by $\gamma_a$, we obtain
\be
  5\im \nabla^a\psi_a = (5\im \gamma^b\nabla_b - \sigma) (\gamma^a\psi_a)\,.
\label{RScon1}
\ee
We should impose the gamma-traceless condition 
\bea
\gamma^a\psi_a=0\,,\label{gammapsi}
\eea
so eqn (\ref{RScon1}) then implies that 
\bea
\nabla^a\psi_a=0\,.\label{divpsi}
\eea

  Now instead act on eqn (\ref{RSop}) with $\nabla_a$.  Using the identity
\be
[\nabla_a,\nabla_b]\psi_c = \ft14 R_{abde}\, \gamma^{de}\,\psi_c +
                          R_{cdab}\, \psi^d\,,\label{commutator}
\ee
it follows after a little algebra that
\be
  \ft{\im}{2}\, R^{ab}\, \gamma_a\,\psi_b= (\sigma- \im \gamma^b\nabla_b)
  (\nabla^a\psi_a)\,.\label{riccicon}
\ee
If the metric is Einstein, this just gives the same conclusion that we
saw before, namely that as well as the gamma-traceless condition 
(\ref{gammapsi}) we shall also have the transversality condition
(\ref{divpsi}).  However, if the metric is not Einstein, we obtain the
independent algebraic condition\footnote{This is the
famous Buchdahl consistency condition 
\cite{Buchdahl:1958xv, Buchdahl:1962, Buchdahl:1982ni}.}
\bea
R_{ab}\, \gamma^a\psi^b=0\,.\label{riccicon2}
\eea

For our present example of the squashed $S^7$ we have in general, 
in vielbein components, 
\be
  R_{\alpha\beta}= a\, \delta_{\alpha\beta}\,,\qquad R_{\hat i \hat j}=
  b\, \delta_{\hat i\hat j}\,,
\ee
where $\alpha=0,1,2,3$ and $\hat i=4,5,6$.  The constants $a$ and $b$ 
depend on the squashing parameter $\lambda$, and are unequal except when
$\lambda^2=1$ or $\lambda^2=\ft15$.  Thus (\ref{riccicon}), together with
the original condition $\gamma^a\,\psi_a=0$, implies that for general
squashing we must have the two independent algebraic conditions
\be
\gamma^\alpha\, \psi_\alpha=0\,,\qquad \gamma^{\hat i}\, \psi_{\hat i}=0\,,
\label{twocon}
\ee
whereas when the metric is Einstein, we have only the condition
$\gamma^a\,\psi_a=0$.  Evidently, then, the Rarita-Schwinger eigenvalue
problem is over-constrained if the metric is not Einstein.

 Since we need to restrict attention to Einstein metrics when 
considering the modes of the Rarita-Schwinger operator, this means that
we must consider only $\lambda=1$ (the round $S^7$) or $\lambda=\ft1{\sqrt5}$
(the Einstein-squashed $S^7$).  It is easily seen that the $SO(8)$ 
representations for the RS modes on the round sphere do not include
any irreps whose decomposition under $Sp(1)\times Sp(2)$ includes any
singlets.  One might be tempted to conclude that there could therefore be
no singlet RS modes in the Einstein-squashed $S^7$.  However, unlike the
situation for all the operators we discussed previously, here we have no
continuous path by which we can deform the RS modes on the round sphere
to the RS modes on the Einstein-squashed sphere.  In fact, as we shall now
show\footnote{This fact was also demonstrated in \cite{Nilsson:2018lof} using 
a group theoretic construction of the irrep spectrum designed for coset spaces.}, there {\it does} exist a singlet RS mode in the Einstein-squashed 
$S^7$.

To see this, we first note that we can construct a class of RS modes in
the Einstein-squashed $S^7$ as follows.  Let $h_{ab}$ be any 
transverse-traceless mode of the Lichnerowicz operator:
\bea
\Delta_L\, h_{ab}= \lambda_L\, h_{ab}\,,\qquad \nabla^a h_{ab}=0\,,\quad
h^a{}_a=0\,.
\eea
we now consider a spin-$\ft32$ field $\psi_a$, defined by
\bea
\psi_a=  h_{ab}\, \gamma^b\eta + \alpha\, (\nabla_c h_{ab})\, \gamma^{cb}\,
\eta\,,\label{psiadef}
\eea
where $\eta$ is the singlet Killing spinor on the Einstein-squashed $S^7$,
obeying eqn (\ref{KSeta}), and $\alpha$ is some constant to be chosen 
later.\footnote{In the
general case, the inclusion of both terms in eqn (\ref{psiadef}) 
is necessary in order that $\psi_a$ can be an eigenfunction of the
Rarita-Schwinger operator.  It will turn out, however, for the
specific singlet mode of $\Delta_L$ that is of principle 
interest to us here, that the two structures in (\ref{psiadef}) are
the same, up to a constant factor.  See later.}
  Given the Killing spinor
equation (\ref{KSeta}) and the fact that $h_{ab}$ is transverse and traceless, 
it is straightforward to see that
$\psi_a$ is divergence-free and gamma-traceless, 
\bea
\nabla^a\psi_a=0\,,\qquad \gamma^a \psi_a=0\,.
\eea
Thus the RS eigenvalue equation (\ref{RSop}) implies that $\psi_a$ obeys
the Dirac eigenvalue equation 
\bea
\im \Dslash_{3/2}\,  \psi_a=\sigma\, \psi_a\,,\label{Diracpsia}
\eea
where $\im\Dslash_{3/2}$ denotes the Dirac operator $\im \gamma^a\nabla_a$ when
acting on vector-spinors.

  Substituting the definition of $\psi_a$ in eqn (\ref{psiadef}) into
eqn (\ref{Diracpsia}) gives equations that 
determine the value that must be chosen for $\alpha$, as
well as determining the eigenvalue $\sigma$.  
After some calculation one finds that eqn (\ref{RSop}) implies
\bea
&&\im\, \Big(-\alpha\, \lambda_L + 7 \alpha\, m^2 +\fft{5\im m}{2}\Big) 
\,h_{ab} \gamma^b\, \eta + \Big(\im + \fft{3\alpha m}{2}\Big)\,
(\nabla_c h_{ab}) \gamma^{cb}\, \eta \nn\\
&&=
\sigma\,\Big[ h_{ab} \gamma^b\eta + \alpha\, (\nabla_c h_{ab}) 
\gamma^{cb}\,\eta\Big]
\eea
The two structures $h_{ab} \gamma^b\eta$ and $(\nabla_c h_{ab}) 
\gamma^{cb}\eta$ are in general linearly independent, and so
equating the coefficients for each implies 
\bea
\sigma = -\im\alpha\,\lambda_L + 7\im \alpha\, m^2 -\fft{5m}{2}=
\fft{\im}{\alpha} + \fft{3m}{2}\,.
\eea
Solving, this gives
\bea
\sigma_\pm &=& -\fft{m}{2} \pm (\lambda_L -3 m^2)^{1/2}\,,\label{sigmaval}\\
\alpha_\pm &=& \fft{2\im m \pm 
   \im \sqrt{\lambda_L-3 m^2}}{(\lambda_L -7m^2)}\,.
\eea
Thus in general a Lichnerowicz mode $h_{ab}$ with eigenvalue $\lambda_L$ 
gives rise to two Rarita-Schwinger modes, with different 
eigenvalues $\sigma_+$ and $\sigma_-$, corresponding to the plus or
minus sign choices.

   If we now apply this construction to the Lichnerowicz  squashing-mode 
 of Page, given in eqn (\ref{Lichmode}), it turns out 
that for this particular case the two structures in the two terms
in the definition of $\psi_a$ in eqn (\ref{psiadef}) 
are proportional to one another, with 
\bea
(\nabla_c h_{ab}) \gamma^{cb} \eta= -\fft{7\im m}{3}\, h_{ab} \gamma^b\eta\,.
\eea
Thus in this case, we may simply choose to write the RS mode in the form
\bea
\psi_a = h_{ab} \gamma^b \eta\,.
\eea
The RS eigenvalue then corresponds just to the minus sign choice in eqn
(\ref{sigmaval}), implying, since $\lambda_L=\fft{28 m^2}{9}$ for the
Lichnerowicz singlet mode, that
\bea
\sigma= -\fft{m}{6}\,.
\eea
This is the eigenvalue of the singlet RS mode on the Einstein-squashed 
$S^7$.

  As discussed above, because there is no way to make a continuous deformation
between modes of the Rarita-Schwinger operator on the round $S^7$ and modes
on the Einstein-squashed $S^7$, it does not necessarily have to be the case
that the RS modes on the Einstein-squashed $S^7$ occur in irreps of
$Sp(1)\times Sp(2)$ that are in 1-1 correspondence with the decomposition
of the $SO(8)$ irreps of the round $S^7$ under the $Sp(1)\times Sp(2)$
subgroup.  We have exhibited a concrete example of this phenomenon in the
construction above.  Namely, we have found that there exists a singlet RS
mode on the Einstein-squashed $S^7$, while on the round $S^7$ none of
the irreps of the RS modes decomposes to include a singlet under
$Sp(1)\times Sp(2)$.  This is very different from the situation for the
singlet Dirac mode on the Einstein-squashed $S^7$ (i.e.~the Killing spinor 
$\eta$), which can be traced back to the round $S^7$ as the singlet in the
decomposition of the 56 Dirac modes in the $(1,0,0,1)$ of $SO(8)$ on the
round sphere.  Unlike the spinor $\eta$, which is responsible for
the space-invader phenomenon in the squashed vacuum, the singlet RS
mode on the Einstein-squashed $S^7$ 
is effectively a ``mode from nowhere,'' which has no counterpart on
the round sphere. This proves the correctness of this particular aspect of 
 the spectrum construction  in \cite{Nilsson:2018lof}.

\section{The Singlets in the AdS$_4$ Spectrum}
  In the previous section, we constructed explicitly the complete set
of singlet modes of all the mass operators that arise in the analysis
of the mass spectrum of the AdS$_4\times M_7$ vacua of eleven-dimensional
supergravity, where $M_7$ is either the round $S^7$ or the Einstein-squashed
$S^7$.  In this section, we shall show explicitly how these give rise to
fields and, in the supersymmetric cases, supermultiplets, in the 
AdS$_4$ vacuum.

    The four-dimensional fields form representations
of the $SO(2,3)$ isometry group of the AdS$_4$ background metric.  They
are denoted by $D(E_0,s)$, where $E_0$ is the lowest energy eigenvalue
and $s$ is the total angular momentum quantum number of the lowest-energy
state.  The representation is unitary if $E_0\ge s+1$ for $s\ge 1$, 
and if $E_0\ge s+\ft12$ for $s=0$ or $s=\ft12$.\footnote{Note that
there exist also the so-called singleton representations
$D(\ft12,0)$ and $D(1,\ft12)$ of $SO(2,3)$, which were discovered
by Dirac \cite{Dirac}.  These will feature in our later discussion.} 
  We refer to 
\cite{heidenreich,dunipo} for further details.
In \cite{dunipo},
the $E_0$ values of the AdS$_4$ representations are listed for each spin: 
\bea
s=0:\qquad \qquad  E_0 &=& \ft32 \pm \ft12 \sqrt{(M/m)^2 + 1}\nn\\
s=\ft12: \qquad\qquad E_0 &=& \ft32 \pm \ft12 |M/m|\nn\\
s=1: \qquad \qquad  E_0 &=& \ft32 +\ft12 \sqrt{(M/m)^2 + 1}\nn\\
s=\ft32: \qquad\qquad E_0 &=& \ft32 + \ft12 |M/m-2|\nn\\
s=2: \qquad \qquad  E_0 &=& \ft32 +\ft12 \sqrt{(M/m)^2 + 9}\,.\label{E0s}
\eea

  It should be noted that in general the eigenvalue spectra of the first-order 
mass operators (Dirac, Rarita-Schwinger and the 3-form operator $Q={*d}$)
will be different if the orientation of the
compactifying seven-manifold $M_7$ is reversed.  Specifically, the
orientation reversal will reverse the signs of all the eigenvalues of
the first-order operators, while the eigenvalues of the second-order 
mass operators
are unaffected.   The
round $S^7$ is an exception to the general rule; the eigenvalues of the
first-order mass operators on the round $S^7$ occur always in pairs, with
every positive eigenvalue having an equal and opposite negative-eigenvalue
partner.  By contrast, the eigenvalue spectra associated with first-order
operators on a squashed $S^7$ are 
asymmetric between positive and negative.  This can be seen in the results
of the previous section, where the totality of singlet modes in the
squashed $S^7$ background is presented, or by consulting the more general results
of \cite{Ekhammar:2021gsg, Karlsson:2021oxd, Karlsson-Nilsson:2023}.

   Because of the asymmetry of the squashed-sphere spectra under orientation
reversal, there are two inequivalent squashed $S^7$ 
compactifications of eleven-dimensional supergravity, as was shown in 
\cite{dunipo0,dunipo}.  These are conventionally referred to as the 
left-squashed compactification and the right-squashed compactification.
The left-squashed vacuum has ${\cal N}=1$ supersymmetry, while the
right-squashed vacuum is non-supersymmetric, so ${\cal N}=0$. 

  For the left-squashed vacuum, we see from the results in section 3,
from the mass operators in eqn (\ref{massops}) and from eqn (\ref{E0s})
that singlets will give rise to the following Heidenreich-type supermultiplets:
The massless supergravity supermultiplet with spin $(2, \ft32)$ and 
two Wess-Zumino multiplets with spin content $(0^+, 0^-, \ft12)$, 
one of which is related to 
singlet modes for the operators $(\Delta_0, Q, i\Dslash_{1/2})$ and the 
second to the modes of the operators
$(\Delta_L, Q, i\Dslash_{3/2})$.

Table \ref{summarytable} summarises the situation for the $Sp(1)\times Sp(2)$ singlet sectors 
of the round $S^7$ vacuum and the left-squashed $S^7$ vacuum. In particular,
one can see where the singlet modes in the left-squashed vacuum have come 
from in the round-sphere vacuum.  Two features in particular are noteworthy:

Firstly, the massless spin-$\ft32$ field (the gravitino) of the 
left-squashed vacuum has not originated from one of the eight massless
gravitini of
the round-sphere vacuum; rather, it has come from the decomposition of
the level $n=1$ $\bf 56_c$ of massive gravitini in the round-sphere
vacuum.  This is the ``space-invaders'' phenomenon that was first 
demonstrated in \cite{dunipo0}. Of course, a corollary of this is that
although the truncation of the full round-sphere spectrum to the singlets
under 
the $Sp(1)\times Sp(2)$ subgroup of $SO(8)$ is a perfectly consistent one,
the singlet modes do not fall into 
supermultiplets in the round-sphere vacuum, because all eight massless gravitini in the round-sphere
vacuum are projected out in this singlet truncation.

Secondly, as we showed in \cite{Nilsson:2018lof} and have been further emphasising in this paper, because the
massless gravitino in the left-squashed vacuum originated from a
massive gravitino in the round-sphere vacuum, there must be an 
``inverse Higgs'', or de-Higgs,  phenomenon, in which the helicity-$\ft12$ states in
the massive gravitino in the round-vacuum are ejected and become a
genuine spin-$\ft12$ mode in the left-squashed vacuum that was not seen
as a distinct spin-$\ft12$ mode in the round-sphere vacuum.  This is
precisely what is seen in the second row of fields listed in
table \ref{summarytable}, where the Rarita-Schwinger mass operator
for the $\ft12^\2$ modes has a singlet mode on the Einstein-squashed $S^7$
but it has no singlet mode on the round $S^7$.  
\bigskip

\renewcommand{\arraystretch}{1.5}

\begin{table}[!ht]
\begin{center}
\begin{tabular}{|c|c||c|c||c|}\hline
\bf Field &\bf Mass operator & \bf Round &\bf Level, $SO(8)$  
 & \bf Left-Squashed \\ \hline\hline
$\ft32^\1$  & $M=-\im\Dslash_{1/2} +\ft{7m}{2}$ &  $M=-m$, $E_0=3$ &$n=1$,
${\bf 56}_c$ &     $M=0$, $E_0=\ft52$       \\ \hline
$\ft12^\2$  & $M= \im\Dslash_{3/2} + \ft{3m}{2}$ & --  & -- &
   $M=\ft{4m}{3}$, $E_0=\ft{13}{6}$   \\ \hline\hline
$\ft12^\4$  & $M=-\im \Dslash_{1/2} - \ft{9m}{2}$ & $M=-9m$, $E_0=6$ &
$n=3$, ${\bf 56}_c$ &
 $M=-8m$, $E_0=\ft{11}{2}$  \\ \hline \hline
$2$ & $M^2 = \Delta_0$ &  $M^2=0$, $E_0=3$ & $n=0$, ${\bf 1}$ & $M^2=0$,
$E_0=3$ \\ \hline
$0^{+\3}$  & $\!M^2\!=
(\sqrt{\Delta_0 \! +\! 9m^2} \! +\! 6m)^2 \!-\! m^2\!
$ &
$M^2=80m^2$, $E_0=6$ &$n=2$, ${\bf 1}$ &  $M^2=80m^2$, $E_0=6$ \\ \hline
$0^{-\2}$ & $M^2=(Q+3m)^2 - m^2$ & $M^2=80m^2$, $E_0=6$ & $n=4$,
${\bf 840}_s$ &
$M^2=48m^2$, $E_0=5$ \\ \hline
$0^{+\2}$  & $M^2=\Delta_L - 4m^2$ & $M^2=24m^2$, $E_0=4$ & $n=2$,
${\bf 300}$ &  $M^2= -\ft{8 m^2}{9}$, $E_0=\ft53$  \\ \hline
$0^{-\1}$  & $M^2= (Q+3m)^2 - m^2$ & $M^2=0$, $E_0=2$ & $n=0$, ${\bf 35}_c$ &
$M^2 = \ft{40 m^2}{9}$, $E_0= \ft83$  \\ \hline
\end{tabular}
\caption{
\label{summarytable}
\it The complete 
spectrum of $Sp(1)\times Sp(2)$ singlets,
in the round and the left-squashed vacua.  The $SO(8)$ reps in the
fourth column contain an $Sp(1)\times Sp(2)$ singlet.  In the squashed
vacuum the ${\cal N}=1$ supermultiplets comprise the fields $(2,\ft32^\1)$,
$(0^{+\3}, 0^{-\2}, \ft12^\4)$ and $(0^{+\2},0^{-\1},\ft12^\2)$.  Note that
the $\ft12^\2$ field occurs only in the squashed vacuum, since
the Rarita-Schwinger operator $\Dslash_{3/2}$ has no $Sp(1)\times Sp(2)$
singlet modes on the round sphere.}
\end{center}
\end{table}

Table \ref{Dgroups} contains a 
complete list of all $Sp(1)\times Sp(2)$ singlets fields in $SO(2,3)$ 
irreps $D(E_0,s)$, grouped into ${\cal N}=1$ supermultiplets in
the supersymmetric left-squashed vacuum.  This table also presents the field
content of the non-supersymmetric right-squashed  vacuum.

We end this section with  a few comments on table \ref{Dgroups} and then in particular on its  relevance  for the 
issue of stability of the right-squashed non-supersymmetric $S^7$ 
compactification.
This theory is well-known to be BF stable and hence,
in view of the AdS stability conjecture proposed in \cite{Ooguri:2016pdq}, 
it is of some interest to see if there 
are other decay modes possible.
One such is connected to the appearance of marginal
operators in the boundary theory \cite{Berkooz:1998qp} (see also, for instance, \cite{Murugan:2016aty}). 
Whether or not the modes analysed in this paper  lead to AdS fields whose dual operators on the boundary 
can be used to construct  such marginal operators depends on the boundary conditions chosen for the 
fields in the third column of the second row 
of table \ref{Dgroups}.  
The fields in the third row cannot give rise to any such operators.
Recall that if a scalar field has $M^2$ in the range $-m^2 \le M^2 \le 3m^2$  then both signs in $E_0=\frac{3}{2}\pm \frac{1}{2}\sqrt{(M/m)^2+1}$  are
compatible with unitarity, the upper sign corresponding 
to Dirichlet and the lower sign to Neumann boundary conditions. 
A similar situation arises for 
Dirac fields with a mass satisfying $|M| \le m$ ($m$ is assumed positive here).

Of the four options appearing in table \ref{Dgroups} for the fields in 
the second row one can for the first case form marginal triple trace operators
dual to three pseudo-scalar fields, $P^3$, or to two fermions and one pseudo-scalar, $P\bar\lambda\lambda$. Here we denote the fields, as well as their dual single trace operators on the boundary,
as $(S, P, \lambda)$. The second option of boundary conditions in table \ref{Dgroups} allows only for a triple trace operator of the kind $P^3$ while the remaining two options cannot be used
for this purpose.

In order to check if these operators  can  give rise to any 
 instabilities one has, however, to compute $1/N$ corrections to the relevant  
 beta-functions in the boundary theory but this is beyond the scope of 
this paper.  

Needless to say it would be quite interesting if one could find arguments that would remove any of the four
possible combinations of boundary conditions from the list on the second row in table \ref{Dgroups}. In fact,
by adopting the point of view advocated in \cite{Nilsson:2018lof} there must be a fermionic singlet mode in the right-squashed case
that has $E_0=1$ and therefore is identified as a singleton.  This 
singleton could be created in another kind of de-Higgsing, in which a
massive AdS$_4$ spin-$\ft12$ fermion in the round-sphere vacuum 
splits up into a singleton and a 
spin-$\ft12$ fermion with a different mass in the right-squashed vacuum. That
this is possible is suggested by the state diagrams in the appendix of
\cite{Nicolai}.  This singleton interpretation 
implies that only the first and the third cases are possible for the
lowest-energies $E_0$ for the $\{0^{+\3},0^{-\2},\ft12^{\4}\}$ fields
in the right-squashed vacuum.

  One further noteworthy feature of the singlet spectrum in the right-squashed
$S^7$ vacuum is that one of the two eigenvalues of the 3-form operator
$Q$ is now $-4m$, which implies that the associated pseudoscalar field
in AdS$_4$ is massless.  This accounts for the lowest-energy possibilities
$E_0=1$ or $E_0=2$ for the $0^{-\2}$ field in the right-squashed vacuum in
table 2.

The rather intriguing corollary is then that while the first case does allow for marginal operators the third case does not.
If the third case 
were to be established as the correct choice of boundary conditions the singlet sector of the theory analysed in this paper would not
lead to any instabilities of the kind discussed here.

\bigskip
\begin{table}[!ht]
\begin{center}
\begin{tabular}{|c|c|c|}\hline
\bf Multiplet (L) & $E_0$ \bf Left-squashed & $E_0$ \bf Right-squashed \\ \hline
$s=\{2,\ft32^\1\}$ & $(3,\ft52)$ & $(3,4)$\\ \hline
$s=\{0^{+\3},0^{-\2},\ft12^{\4}\}$& $(6,5,\ft{11}{2})$  &$(6,1,1)$ 
 or $(6,1,2)$ or $(6,2,1)$ or $(6,2,2)$ \\ \hline
$s=\{0^{+\2},0^{-\1},\ft12^\2\}$& $(\ft53, \ft83, \ft{13}{6})$ &
 $(\ft43,\ft{10}{3},\ft73)$ or $(\ft53, \ft{10}3, \ft{7}{3})$\\
 \hline
\end{tabular}
\caption{
\label{Dgroups}
 \it The modes in the multiplets are grouped
according to the ${\cal N}=1$
supersymmetry of the left-squashed vacuum.  The groupings for the
right-squashed vacuum represent the same $S^7$ modes as in the left-squashed
vacuum, now with sign-reversals for the $\im \Dslash_{1/2}$,
$\im \Dslash_{3/2}$ and $Q$ eigenvalues.  The alternative possibilities
for the $E_0$ values in 
the right-squashed vacuum represent the sign ambiguities for $E_0$ for
spins 0 and $\ft12$; these are discussed in the text.}
\end{center}
\end{table}

\section{Conclusions}

In this paper we considered the seven-sphere compactification of 
eleven-dimensional supergravity,
and discussed the relation between  the theories in AdS$_4$ resulting from 
using the round $S^7$  and its two squashed versions, 
the left-squashed $S^7$ that gives ${\cal N}=1$ supersymmetry
 and the orientation-reversed right-squashed $S^7$, which gives
no supersymmetry. 

Our discussion relied on the explicit construction of all the generalised
Fourier 
modes of the relevant operators on $S^7$ 
that are singlets under the squashed $S^7$ isometry group $Sp(1)\times Sp(2)$. 
These modes were then analysed with respect to their properties under
the continuous homogeneous deformation from the round to the 
Einstein-squashed $S^7$. While we showed that the singlet modes of 
the bosonic operators and the spin-$\ft12$ Dirac operator 
can indeed be followed in this way, this is not the case for the 
Rarita-Schwinger singlet mode  on the squashed $S^7$. 
We then related this to a 
special feature for higher-spin operators first discussed  by  
Buchdahl  \cite{Buchdahl:1958xv, Buchdahl:1962, Buchdahl:1982ni}.  
This feature of the compactification is, however, in full accord with 
the squashed spectrum  constructed in \cite{Nilsson:2018lof},
which clearly shows that the Rarita-Schwinger singlet mode does not 
emanate from any mode in the round sphere operator spectrum.

In the  AdS$_4$  theory coming from the left-squashed $S^7$ we may 
also identify the $SO(2,3)$ ${\mathcal N}=1$ supersymmetry multiplets 
associated with the singlet modes.
These are the 
${\mathcal N}=1$ supergravity multiplet and two Wess-Zumino multiplets. 
One of the latter is the so called ``Page multiplet,'' which 
contains a scalar field that was shown by Page
to be responsible for the squashing of the $S^7$. Its origin in the round 
sphere spectrum was also given. Curiously enough, as we explained here, 
the spin-$\ft12$ field in this multiplet is associated with a Rarita-Schwinger
mode in the Einstein-squashed $S^7$ that has no 
origin in the round sphere. Instead, from the point of view of the 
AdS$_4$ theory, this spin-$\ft12$ field is ejected from a massive 
spin-$\ft32$ field in the round sphere vacuum, when
it becomes massless on the left-squashed sphere. In other words,
this is a kind of inverse Higgsing, or de-Higgsing. This particular 
point  was also emphasised in \cite{Nilsson:2018lof}.

Under an orientation-reversal of the squashed $S^7$, the ${\cal N}=1$  
supersymmetry of the left-squashed vacuum is lost, which seems to 
introduce certain ambiguities concerning the 
boundary conditions to be imposed on the different scalar and fermionic fields. For some of the fields it is 
not clear how to discriminate between Dirichlet and Neumann boundary 
conditions. In the left-squashed  case this issue is easily resolved 
by appealing to  supersymmetry; the singlet fields  must necessarily 
fall into ${\cal N}=1$ supermultiplets, but there is no such
requirement in the right-squashed case.
In fact, although the operator eigenvalues can be followed as a function of the 
squashing parameter this is probably not the case for the masses since their 
relation to the operators are not only theory dependent but also valid only 
in backgrounds that solve the field equations. This also presents a
difficulty for 
following the change in $E_0$ as a function of the squashing parameter.

We have also discussed the relevance of our results for the 
issue of stability of the right-squashed non-supersymmetric $S^7$ 
compactification.
This theory is BF stable but we found in the previous section
that one can form  marginal
operators in the boundary theory. 
As is clear from  table 2 above, this depends on the boundary conditions that are used 
for some of the singlet fields after skew-whiffing, where some options   
make it possible to construct marginal triple-trace operator. However, as mentioned at the end of the
previous section, by adopting the reasoning of  \cite{Nilsson:2018lof}, there must exist a fermionic 
singleton in the right-squashed vacuum, and
this excludes two of the four options on the second row of table \ref{Dgroups}.
One of the two options that remain can then not lead to any marginal operators and would provide a theory that 
could still be completely stable.

 In order to check if the marginal  operators do indeed  generate
 instabilities one has, however, to compute $1/N$ corrections to 
some beta-functions in the boundary theory, which is beyond the scope of 
the present paper.  This would, however, be an interesting topic for a future
investigation.

\section*{Acknowledgments}
We are grateful to M.J. Duff for extensive discussions. B.E.W.N. 
thanks also Joel Karlsson for
discussions related to this work. The work of C.N.P. 
is supported in part by DOE grant DE-SC0010813.  B.E.W.N. is 
partly supported by
the Wilhelm and Martina Lundgren Foundation.

\end{document}